\documentclass[a4paper,twoside]{article}

\usepackage{epsfig}
\usepackage{subcaption}
\usepackage{calc}
\usepackage{amssymb}
\usepackage{amstext}
\usepackage{amsmath}
\usepackage{amsthm}
\usepackage{multicol}
\usepackage{pslatex}
\usepackage{apalike}
\usepackage{algorithm2e}
\usepackage[bottom]{footmisc}
\usepackage{graphicx}
\usepackage{SCITEPRESS}     % Please add other packages that you may need BEFORE the SCITEPRESS.sty package.

\begin{document}

\title{Guardian Crawler: Retrieval-First Knowledge Discovery with Bounded LLM Augmentation for Noisy Web Intelligence}

% \author{\authorname{First Author Name\sup{1}\orcidAuthor{0000-0000-0000-0000}, Second Author Name\sup{1}\orcidAuthor{0000-0000-0000-0000} and Third Author Name\sup{2}\orcidAuthor{0000-0000-0000-0000}}
% \affiliation{\sup{1}Institute of Problem Solving, XYZ University, My Street, MyTown, MyCountry}
% \affiliation{\sup{2}Department of Computing, Main University, MySecondTown, MyCountry}
% \email{\{first\_author, second\_author\}@ips.xyz.edu, third\_author@dc.mu.edu}
% }
\author{\authorname{Joshua Castillo\orcidAuthor{0009-0006-2660-5448}, Santosh Nukavarapu\orcidAuthor{0000-0001-7922-7539}, and Ravi Mukkamala\orcidAuthor{0000-0001-6323-9789}}
 \affiliation{Department of Computer Science}
 \affiliation{Old Dominion University}
 \email{\{jcast046, snukavar, rmukkama\}@odu.edu}
 }

% Information Retrieval, Knowledge Discovery, Web Intelligence, LLM-Grounded Summarization, Trustworthy AI, Missing-person Intelligence.}

\abstract{Retrieving relevant evidence from noisy web data is challenging, particularly in sensitive domains containing incomplete reports, heterogeneous language, and irrelevant content. We present Guardian Crawler, a reproducible retrieval-first testbed for controlled experiments on knowledge discovery and evidence-grounded summarization over synthetic web-like corpora. The architecture combines BM25 retrieval with risk-aware, embedding-augmented, and hybrid reranking, followed by constrained retrieval-augmented generation with explicit document citations. Experiments on a synthetic 900-document corpus and 10 queries produced the highest descriptive retrieval scores under risk-based reranking, with P@10 = 1.00 and NDCG@10 = 0.94, compared with 0.94 and 0.81 for BM25. The best hybrid and BM25+Semantic configurations reached NDCG@10 values of 0.94 and 0.88, respectively. All 41 evaluable generated bullets passed the lexical coverage threshold; an automated LLM judge classified 36 as supported, one as partially supported, and four as unsupported. These results demonstrate the feasibility of Guardian Crawler as a controlled testbed but do not establish statistical superiority, human-validated faithfulness, or transfer to live-web investigative environments.
}

% Collectively, these findings underscore the utility of synthetic evaluation environments, risk-aware retrieval methods, and tightly grounded summarization for evidence-sensitive web search scenarios.}

\onecolumn \maketitle \normalsize \setcounter{footnote}{0} \vfill

\section{\uppercase{Introduction}}
\label{sec:introduction}
Web-based knowledge discovery requires relevant evidence to be collected, normalized, ranked, and evaluated before downstream analysis can be considered reliable. This is particularly important in sensitive domains, where relevant information may be distributed across official bulletins, news reports, reposts, incomplete descriptions, and unrelated content. Ranking or synthesis errors can obscure important evidence or produce unsupported conclusions (Croft et al., 2015; Manning et al., 2009).

Missing-child and trafficking-related information provides a demanding setting for studying this problem, but unrestricted experimentation on live data creates substantial ethical and privacy concerns. Guardian Crawler therefore uses a synthetic, link-connected mini-web that emulates bulletin, news, noise, and irrelevant pages while preserving control over corpus composition, metadata, and relevance rules.

Guardian Crawler is a controlled experimental platform rather than a deployed investigative system. It uses BM25 as a transparent lexical baseline and limits large language models to bounded roles in corpus construction, document tagging, citation-bearing summary generation, and automated grounding assessment. Risk-aware and embedding-based signals are applied as rerankers over a shared candidate pool rather than replacing the retrieval pipeline with an end-to-end generative model.

The study addresses three questions: whether a synthetic mini-web can support reproducible retrieval experiments; how risk-aware, embedding-augmented, and hybrid reranking behave relative to BM25 under fixed conditions; and whether constrained generation can produce citation-bearing summaries whose support can be evaluated separately from retrieval quality.

Across 10 queries and 900 documents, risk-based reranking produced the highest observed P@10 and NDCG@10 values, while BM25+Semantic reranking produced the strongest semantic-alignment diagnostic. Automated grounding evaluation also revealed a difference between lexical coverage and judge-assessed support. These are descriptive results from a small, domain-specific synthetic benchmark and should not be interpreted as statistically supported evidence of general retrieval superiority or operational effectiveness.

The paper contributes: (1) a reproducible synthetic testbed for controlled studies of noisy web-like retrieval; (2) a modular retrieval, reranking, generation, and evaluation pipeline with bounded LLM roles; and (3) a comparative analysis that separates retrieval effectiveness from automated grounding assessment.
\section{\uppercase{Related Work}}
Guardian Crawler builds on established information-retrieval architectures that separate acquisition, indexing, ranking, and evaluation rather than treating search as an opaque end-to-end process \cite{croft2015search,manning2009ir}. Research on web mining and unstructured-document analysis similarly emphasizes the importance of robust normalization and retrieval before higher-level inference is attempted \cite{aggarwal2015data,bielska2020osint,mahadevkar2024ai}.

Retrieval-augmented generation reduces reliance on parametric memory by conditioning generation on external evidence \cite{chen2022retrieval}. Guardian Crawler applies this principle narrowly: generation occurs only after retrieval, uses a bounded evidence set, requires document identifiers, and is evaluated separately for grounding. The system therefore resembles evidence-grounded summarization more than general-purpose question answering.

LLMs can also serve as constrained annotators or evaluators, although their reliability depends on task specification, structured outputs, and independent validation \cite{chen2024llmannotator,ratner2017snorkel}. Guardian Crawler uses schema-restricted prompts but does not treat LLM outputs as ground truth. The current study evaluates embedding similarity only as a second-stage reranking signal; it does not benchmark an independent first-stage dense retriever or an external neural retrieval system. Its empirical comparisons are therefore limited to ranking formulations implemented within the Guardian architecture.

\section{\uppercase {System Design and Architecture}}
Three design goals shape the Guardian Crawler system. The first is reproducibility: experiments should be rerunnable over a controlled corpus with known document types, link structure, and relevance heuristics. The second is modularity: retrieval, generation, tagging, embeddings, and evaluation should be independently inspectable and replaceable. The third is boundedness: LLMs should be used where they plausibly add value, but not as opaque substitutes for crawling, indexing, or ranking. These goals align with conventional search-engine engineering wisdom while also responding to contemporary concerns about LLM overreach \cite{croft2015search,floridi2019ai}.

\begin{figure}[h]
  \centering
  \includegraphics[width=\linewidth]{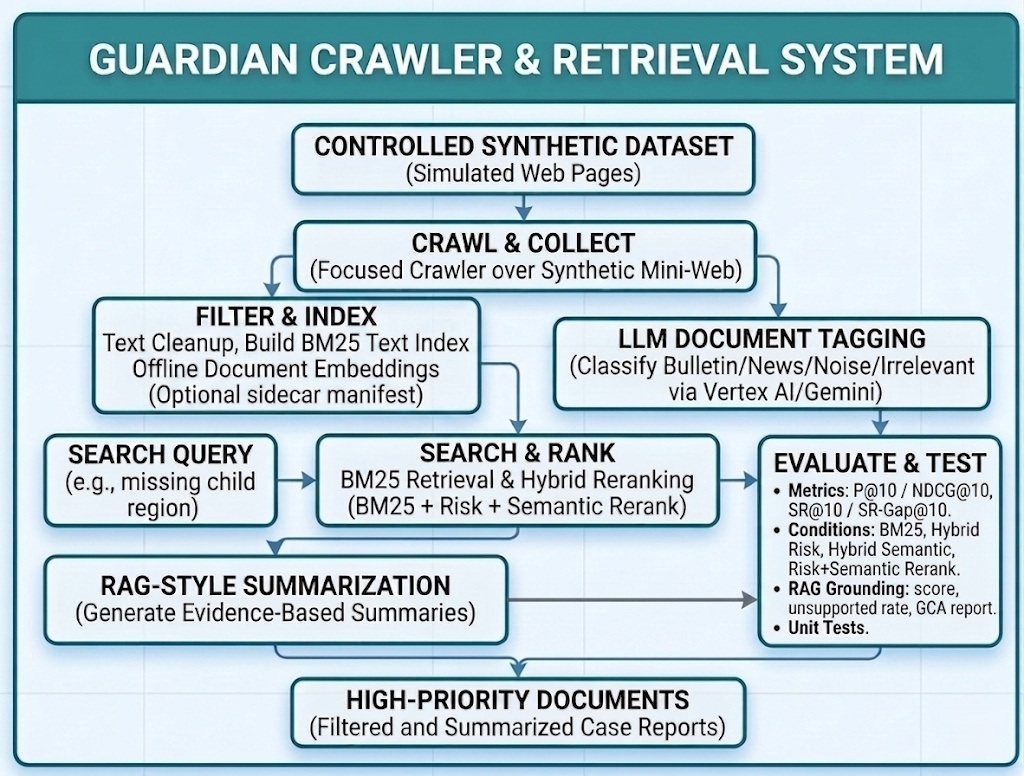}
  \caption{Guardian Crawler Architecture}
  \label{fig:Arch}
\end{figure}

The system architecture is organized as an end-to-end but modular pipeline (Figure~\ref{fig:Arch}). At a high level, the configuration layer sets shared defaults; the mini-web generator creates the corpus; the HTML extractor standardizes raw pages; the crawler collects them; the BM25 index provides first-stage retrieval; the tagging layer assigns coarse page types; the risk reranker promotes harm-relevant lexical signals; the semantic reranker uses embeddings to refine BM25 candidates; the ranking runner compares all conditions fairly; the RAG layer converts retrieved evidence into citation-linked bullets; and the evaluation layer scores both retrieval and grounding. We  now describe each component in detail. 

\subsection{Synthetic Corpus and Acquisition}
The mini-web generator creates a static HTML corpus from structured case seeds and assigns pages to four high-level document types: bulletin, news, noise, and irrelevant. Hub pages and synthetic hyperlinks produce a crawlable graph with same-type linking bias and enforced reachability. Using generated rather than live content preserves experimental control and reduces privacy risk while retaining retrieval challenges such as mixed relevance, redundancy, and distracting content.

A breadth-first crawler operates within the synthetic site, normalizes URLs, enforces depth and page limits, and removes duplicate content through hashing. A shared HTML extraction component produces a consistent representation of each page’s title, visible text, and outbound links for crawling and indexing.

\subsection{Retrieval and Reranking}
BM25 indexes normalized title and body text and serves as the common lexical baseline. Risk-aware reranking combines normalized BM25 scores with a bounded score derived from a 28-entry harm-related lexicon. Embedding-based reranking computes similarity between query embeddings and precomputed document embeddings. Hybrid conditions combine lexical, risk, and semantic signals.

All ranking conditions operate on the same BM25-derived candidate set. In the reported experiment, the candidate pool contains all 900 corpus documents. The comparison therefore isolates scoring and reranking behavior rather than first-stage candidate recall or retrieval efficiency. The embedding component is a reranker and not an independent dense-retrieval baseline; every reported BM25+Semantic condition retains a nonzero BM25 contribution.

\subsection{Bounded LLM Components}
Gemini is used in four separate roles: synthetic corpus construction, post-crawl page tagging, citation-bearing summary generation, and automated grounding assessment. Tagging maps each page to bulletin, news, noise, or irrelevant using a fixed JSON schema. The generation stage receives only selected evidence passages and must return a limited set of bullets with supporting document identifiers. Outputs may abstain when evidence is insufficient or validation fails.

The judge receives the query, generated bullet, and cited evidence and assigns supported, partially supported, or unsupported labels with an issue category. Separate generator and judge instances provide procedural separation, but they do not provide model-family independence because Gemini participates in multiple stages.

\subsection{Evaluation Architecture}
Retrieval effectiveness and grounding are evaluated separately. Retrieval qrels are generated from synthetic metadata and query-specific relevance rules. Grounding is assessed first through deterministic lexical coverage and then through the LLM judge. Neither process constitutes human validation: qrels are rule-derived, and grounding labels are generated automatically.

\section{\uppercase{Experimental Setup}}
The canonical experimental configuration employs a synthetic corpus comprising 900 documents, a candidate pool of 900 documents per query, and a set of 10 queries. The comparison framework includes a BM25 baseline, four BM25+risk configurations with $\alpha \in \lbrace 0.25, 0.5, 0.75, 1.0\rbrace$, three BM25+semantic configurations with $\lambda \in \lbrace 0.2, 0.4, 0.6\rbrace$, and three hybrid lexical–risk–semantic configurations with weight triplets (0.6, 0.2, 0.2), (0.5, 0.25, 0.25), and (0.4, 0.3, 0.3). All configurations are evaluated under a unified reranking protocol, ensuring that each condition operates on the same BM25-derived candidate pool for every query.
\subsection{Implementation details and reproducibility}
Retrieval is performed over a closed synthetic corpus, and BM25 is the only first-stage retriever. All risk-aware, semantic, and hybrid conditions rerank a fixed BM25 candidate list, ensuring that later comparisons reflect scoring differences rather than different recall frontiers. Query terms are treated as a set rather than a multiset during BM25 scoring. 

The BM25 implementation uses default parameters $k_1=1.2, b = 0.75$, and $min\_doc\_tokens = 3$. IDF (Inverse Document Frequency) is computed as $\log \frac{(N-df+0.5)}{(df+0.5)}+1$,where $N$ is the total number of documents in the corpus (900) and $df$ is the document frequency or the number of documents that contain a given term. It sums the per-term BM25 contributions over distinct query tokens. Tokenization lowercases the text and extracts alphanumeric tokens using the regular expression [a-z0-9]+.

The risk-aware reranker uses a weighted lexicon with 28 entries. For each document d, raw risk is computed over title and body as
 $RiskRaw(d)=\sum_t w_t min(\frac{tf_t(d)}{2})$,
 where each term match count is capped at 2. The bounded risk score is then
$RiskScore(d) = \frac{RiskRaw(d))}{RiskRaw(d)+3}$

 % BM25 scores are min-max normalized within the candidate pool, whereas RiskScore is already bounded in [0,1]. The risk reranker uses
 % $(1-\alpha)*BM25+\alpha*RiskScore$,
 % and the triple hybrid with weights $(w_B, w_R, w_S)$ uses
 % $w_B*BM25+w_R*RiskScore+w_S*Sem(d,q)$, 
 % with semantic similarities also min-max normalized on the same candidate set. Here, $Sem(d,q)$ represents the embedding-based semantic similarity between the query (q) and document (d), usually computed from the query embedding and the precomputed document embedding within the fixed BM25 candidate pool.
 $\widehat {BM25}$ scores are min-max normalized within the candidate pool, whereas RiskScore is already bounded in [0,1]. The risk reranker uses the following three scores:
    $$Score_{risk}(d,q)=(1-\alpha)*\widehat{BM25}(d,q)+\alpha*Risk(d)$$
    $$Score_{sem}(d,q)=(1-\lambda)*\widehat{BM25}(d,q)+\lambda*\widehat{Sem}(d,q)$$
    $$Score_{hybrid}(d,q)=w_B*\widehat{BM25}(d,q)+w_R*Risk(d)$$
    $$+w_S*\widehat{Sem}(d,q)$$
where $w_B, w_R,$ and $w_S$ are the lexical, risk, and semantic weights, respectively, and $\widehat{Sem}(d,q)$ is the normalized embedding-based query-document similarity.

Document embeddings are built offline from the BM25 index rather than directly from raw crawl files. For each doc\_id, the system embeds a canonicalized title-plus-body text representation, fingerprints that representation, and writes both JSONL vectors and a manifest containing the model, task types, dimensionality, maximum character budget, canonicalization version, and similarity mode. 

% At rerank time, the sidecar manifest is validated against the live run, and each candidate document’s stored content fingerprint must match a recomputed fingerprint from the current index. Any mismatch causes the semantic reranker to fail fast rather than silently proceed with stale vectors. Default model settings are VERTEX\_MODEL = gemini-2.5-flash, VERTEX\_FALLBACK\_MODEL = gemini-2.5-pro, and EMBEDDING\_MODEL = text-embedding-005. 
\subsection{Ranking conditions and candidate-pool control}
The four compared ranking conditions are:
\begin{enumerate}
 \item BM25 as the baseline lexical retriever;
 \item BM25 + Risk at $\alpha \in \lbrace 0.25, 0.5, 0.75, 1.0\rbrace$;
 \item BM25 + Semantic at $\lambda \in \lbrace 0.2, 0.4, 0.6\rbrace$; and
 \item BM25 + Risk + Semantic with weight triplets: (0.6, 0.2, 0.2), (0.5, 0.25, 0.25), and (0.4, 0.3, 0.3). 
 \end{enumerate}
The candidate set is fixed to the top 900 BM25 retrieval results, which in this experiment coincides with the full corpus. This shared-candidate design is critical, as it isolates reranking behavior by keeping the underlying evidence pool constant across all experimental conditions.

Semantic reranking is implemented strictly as a second-stage procedure. It does not replace BM25 as the initial retrieval step; instead, it only reorders the BM25-derived candidate set based on dense similarity between the query embedding and the corresponding precomputed document embeddings. Hybrid ranking similarly integrates lexical, risk-sensitive, and semantic signals over this same candidate list. Consequently, the BM25 configuration serves as the common reference baseline for all subsequent comparative analyses.

\subsection{Retrieval metrics and query relevance judgments (qrel) construction}
Retrieval effectiveness is evaluated primarily using P@10 and NDCG@10, which measure the proportion of relevant documents in the top 10 and rank-sensitive graded retrieval quality, respectively. When embedding-based representations are enabled, the system additionally reports SR@10 and SR-Gap@10, the semantic closeness of relevant top-10 results to the query  and how much more semantically similar the relevant top-10 documents are as compared to non-relevant ones, respectively. These two metrics quantify the semantic alignment between queries and the top-ranked documents under the embedding model. SR@10 is defined as the qrel-weighted mean similarity over relevant documents appearing in the top ranks, whereas SR-Gap@10 captures the difference in similarity scores between relevant and non-relevant documents among the top-ranked set. These embedding-aware measures are intended as diagnostic indicators rather than substitutes for standard ranked-retrieval metrics.

Relevance judgments are instantiated as rule-derived qrels rather than pooled human assessments. Relevance grades take values in $\lbrace0, 1, 2\rbrace$ and are generated from metadata.jsonl and query\_relevance\_rules.json. 
For example, in query set Q1, bulletin and news pages are assigned a base relevance grade of 2, non-relevant pages receive a grade of 0, and “noise” pages receive a grade of 1 unless they contain at least two predefined cue terms (e.g., “missing,” “child,” “last seen”) in the concatenated page text, in which case the page’s relevance is boosted according to the specified query rule. This closed-world labeling procedure yields complete query–document relevance annotations for the synthetic corpus and ensures exact repeatability, but it is not directly comparable to large-scale human relevance assessment. Because both the qrel rules and the risk lexicon encode related domain cues, their unmeasured overlap may partially favor risk-aware ranking. The resulting qrels were not validated through human relevance assessment.
\subsection{Grounding evaluation protocol}
Grounded generation is evaluated in two stages. 
First, a deterministic lexical support check measures whether the content tokens in each cited bullet are covered by the concatenated cited evidence. Let $C_b$ and $C_e$ denote token-multiset counts for the bullet and its cited evidence, respectively. Token coverage is defined as
 $Cov(b,e)= \frac{\sum_t min(C_b(t),C_e(t))}{\sum_t C_b (t)}$.
 A bullet is counted as lexically supported if $Cov(b,e)\geq\tau$, where the default threshold is $\tau$ = 0.35. Only evaluable cited bullets are included: bullets must contain non-empty text and non-empty supporting\_doc\_ids. Rows marked with failure\_stage, abstained, or lacking evaluable cited bullets are excluded from lexical support aggregates.

Second, an optional LLM-as-judge protocol evaluates support more semantically. The judge sees the query, the bullet text, and cited evidence only, and returns a structured JSON label with supported, partially\_supported, or unsupported, together with a confidence score, an issue\_type, and a short reason. The issue taxonomy includes none, overstatement, citation\_mismatch, not\_in\_evidence, ambiguous, and multi\_hop\_inference. Judge aggregation reports the strict judge-assessed support rate (supported only), lenient judge-assessed support rate (supported + partially supported), unsupported claim rate, and issue histograms. The judge is run separately from the generator and cached on disk for reproducible reruns. These labels are automated model assessments and were not validated by human annotators.
\subsection{Reproducibility and statistical scope}
\begin{figure}[h]
  \centering
  \includegraphics[width=\linewidth]{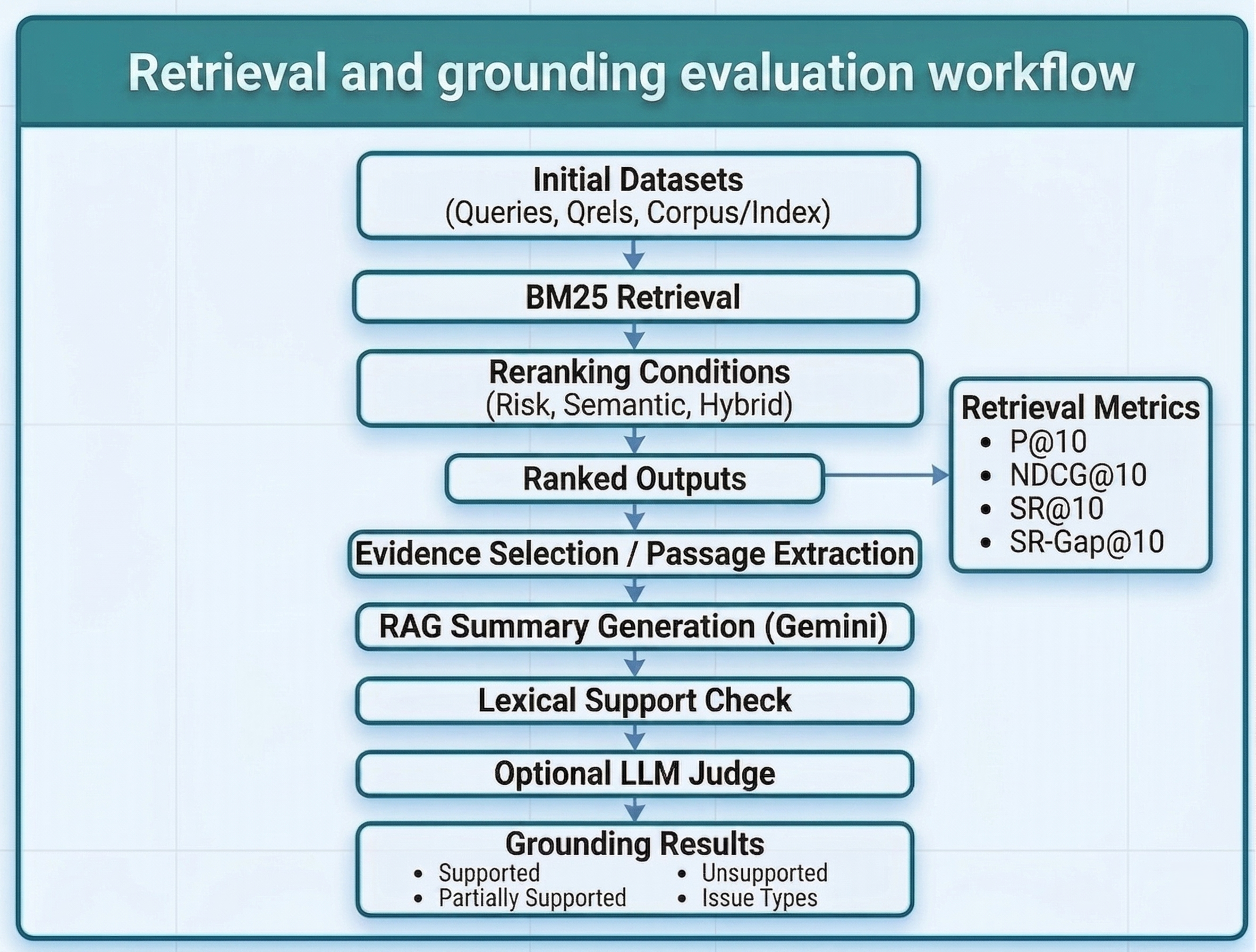}
  \caption{Retrieval and grounding evaluation workflow}
  \label{fig:fig2}
\end{figure}

A unified evaluation pipeline regenerates the qrels, reruns the ranking comparisons, and computes retrieval and grounding summaries, ensuring that all reported results originate from the same auditable workflow illustrated in Figure~\ref{fig:fig2}.
%The end-to-end experiment is coordinated via a unified evaluation pipeline that regenerates relevance judgments (qrels), re-executes the ranking comparisons, formats retrieval summaries, and computes scores for RAG grounding reports. This integrated orchestration is critical because it ensures that all reported results derive from a single, auditable workflow, rather than from manually aggregated outputs produced by heterogeneous and potentially inconsistent scripts. The information workflow is illustrated in Figure~\ref{fig:fig2}.

The experiment does not include paired significance tests, bootstrap-based confidence intervals, or permutation tests for either retrieval or grounding metrics. Consequently, the macro-level differences reported in Section 5 should be viewed as comparative outcomes on a controlled benchmark, rather than as statistically supported claims of general superiority beyond the specific experimental conditions considered here. This limitation is especially salient because the evaluation set comprises only 10 queries, which is adequate for structured ablation studies but small relative to standard information retrieval benchmarking practice \cite{croft2015search,manning2009ir}. %Figure 2 summarizes the retrieval and grounding evaluation workflow, from qrels and BM25 retrieval through reranking, evidence selection, RAG generation, lexical support checking, optional LLM judging, and final grounding analysis.
\subsection{Representative pipeline example}
A representative end-to-end example is query Q1: “missing child last seen near bus stop.” Under the BM25 baseline, the highest-ranked documents are, in order, D0519, D0139, D0789, D0691, and D0781. Under BM25+Risk with $\alpha = 1.0$, the ranking changes substantially: D0789 becomes the top-ranked document, followed by D0402, D0589, D0837, and D0677, while D0519, originally ranked first under BM25, falls out of the top-10 results. In this example, risk-aware reranking does not merely resolve near ties but substantially reorders the candidate set when harm-relevant lexical cues are prominent.

For the same query, the grounded generation layer produces the following bullet: “A 13-year-old male named Jordan Lee is missing from Richmond, VA, and was last seen near I-95 on January 15, 2025, wearing a red hoodie and Spider-Man flip-flops. A gold 2015 Toyota Sienna minivan is connected to his disappearance.” The cited supporting document is D0519, and the corresponding evidence passage explicitly includes the name, age, location, clothing description, and vehicle reference. In the lexical grounding stage, this bullet is classified as supported under the token-coverage criterion. In the LLM-judge stage, it is labeled as supported with confidence 1.0 and issue type “none.” This example illustrates the intended success mode of the system: lexical retrieval surfaces the relevant evidence, grounded generation produces a concise summary, and both lexical and judge-based evaluation concur that the resulting claim is supported.

\section{\uppercase{Results}}
A summary of the results obtained during our experiments using Guardian Crawler with a corpus of 900 documents is shown in Table~\ref{results}. We now provide our observations on the results.
\begin{table}
\centering
\caption{Macro retrieval results on the controlled 900-document, 10-query synthetic benchmark. BM25 is augmented with Risk (R) and/or Semantic (S) reranking. The metrics are P@10 (M1), NDCG@10 (M2), SR@10 (M3), and SR-Gap@10 (M4). Results are descriptive; no statistical significance testing was performed.}
\label{results}
\footnotesize
\begin{tabular}{|l|l|l|l|l|}
\hline
Model&M1&M2&M3&M4\\
\hline
BM25&0.94&0.81&0.55&0.14\\
BM25+R($\alpha=0.25)$&0.92&0.81&0.55&0.16\\
BM25+R($\alpha=0.50)$&0.97&0.90&0.55&0.12\\
BM25+R($\alpha=0.75)$&1.00&0.93&0.55&--\\
BM25+R($\alpha=1.0)$&1.00&0.94&0.54&--\\
BM25+S($\lambda=0.2)$&0.94&0.83&0.56&0.16\\
BM25+S($\lambda=0.4)$&0.97&0.87&0.57&0.04\\
BM25+S($\lambda=0.6)$&0.97&0.88&0.57&0.03\\
BM25+R+S($0.6,0.2,0.2)$&0.94&0.83&0.55&0.17\\
BM25+R+S($0.5,0.25,0.25)$&0.96&0.88&0.55&0.04\\
BM25+R+S($0.4,0.3,0.3)$&0.99&0.94&0.56&0.01\\
\hline
\end{tabular}
\end{table}

\subsection{Retrieval effectiveness}
On this controlled benchmark, BM25+Risk at $\alpha$ = 1.0 produced the highest observed retrieval scores, with P@10 = 1.00 and NDCG@10 = 0.94. The BM25 baseline yielded (0.94, 0.81), the best hybrid condition with weights (0.4, 0.3, 0.3) yielded (0.99, 0.94), and the best BM25+Semantic condition at $\lambda$ = 0.6 yielded (0.97, 0.88).
%The principal retrieval result is clear: the strongest overall ranking condition is BM25+Risk at $\alpha$ = 1.0, with P@10 = 1.00 and NDCG@10 = 0.94, represented as the pair (1.0,0.94). The BM25 baseline yields (0.94, 0.81). The best hybrid condition, BM25+risk+semantic with weights (0.4, 0.3, 0.3) yields (0.99, 0.94). The best semantic-only condition, BM25+Semantic at $\lambda$ = 0.6 yields (0.97, 0.88).

These results suggest three benchmark-specific observations. First, risk-aware reranking produced the highest macro retrieval scores in this setting. Second, BM25+Semantic reranking improved over BM25 but did not exceed the highest risk-aware scores on the primary metrics. Third, the best hybrid condition approached the highest risk-aware NDCG@10, suggesting that semantic signals provided complementary value within this corpus.

The $\alpha$ sweep shows that stronger risk weighting generally produced higher NDCG@10 after $\alpha$ = 0.25, culminating in the highest observed value at $\alpha$ = 1.0. Because the evaluation contains only 10 queries and uses rule-derived qrels, this should be interpreted as a descriptive result specific to the present benchmark rather than evidence of general superiority.
%The $\alpha$ sweep reinforces this pattern. Risk-aware reranking improves steadily from weaker to stronger risk weighting, culminating in the best NDCG at $\alpha$ = 1.0 in the authoritative metrics summary. This suggests that, for the evaluated queries, the domain lexicon is aligned closely enough with the relevance structure that aggressive weighting helps rather than harms. That is not a universal property of reranking systems, but it is an important empirical finding here.

\subsection{Lexical relevance, risk sensitivity, and semantic alignment}
The observed performance of risk-aware reranking may reflect the characteristics of the information needs represented in the benchmark. Risk-aware reranking appeared most useful for queries containing explicit warning-style formulations, whereas BM25+Semantic reranking may provide complementary value for paraphrastic or indirectly expressed requests. Put differently, within the corpus there exist tasks for which the decisive criterion among relevant documents is not solely topical alignment, but also the intensity and specificity of harm-related language. BM25 is able to retrieve a broad set of relevant pages; risk-aware reranking then more effectively prioritizes those documents that are most consequential under such query formulations.

Semantic reranking, by contrast, presents a complementary picture. Its best macro NDCG is lower than the best risk-aware configuration, yet its semantic evaluation indicators are robust. In the aggregate metrics summary, the strongest BM25+Semantic setting ($\lambda = 0.6$) achieves the highest SR@10 of 0.57 among the compared conditions, exceeding both BM25 and BM25+Risk. In this benchmark, BM25+Semantic reranking was associated with higher semantic proximity even when it did not maximize the primary retrieval metrics. %This distinction is non-trivial: semantic reranking can enhance the semantic proximity of retrieved relevant documents to the query even when it does not maximize the primary macro retrieval objective.

Such a trade-off is consistent with a retrieval-first architecture. Within this benchmark, lexical and risk-aware signals were effective when queries contained explicit markers aligned with the domain-specific vocabulary. Embedding-based representations may provide complementary value when information needs are expressed indirectly through paraphrase, implication, or looser semantic relatedness. The results therefore suggest that lexical, risk-aware, and embedding-based signals addressed partially different ranking characteristics within this benchmark.

\subsection{Grounded generation and faithfulness}
The RAG layer yields promising, though not flawless, grounding performance. The deterministic lexical support procedure identifies 41 of 41 evaluable bullets as supported, corresponding to 100\% lexical support. In contrast, the LLM-as-judge assessment applies a more stringent criterion, reporting an 87.8\% strict judge-assessed support rate, a 90.2\% lenient judge-assessed support rate, and a 9.8\% unsupported-claim rate across the same 41 bullets. The majority of identified issues are attributed to multi-hop inference, with a smaller subset arising from not-in-evidence errors.

The discrepancy between lexical support and judge-based support is informative. Within this automated evaluation, it shows that lexical overlap alone is insufficient to establish faithfulness. A bullet may reuse extensive lexical material from the cited passages yet still over-aggregate across sources, conflate distinct cases, or introduce inferences that are not warranted by the underlying evidence. The design of Guardian Crawler surfaces this distinction explicitly rather than collapsing it into a single scalar “RAG quality” metric. Because no human adjudication was performed, these rates should be interpreted as automated judge assessments rather than human-validated faithfulness estimates.

\subsection{Overall interpretation}
Taken together, the results support the feasibility of Guardian Crawler as a controlled experimental testbed. Within this benchmark, risk-aware reranking produced the highest descriptive top-rank scores, BM25+Semantic reranking improved semantic-alignment diagnostics, and grounded generation remained vulnerable to evidence-merging errors. These findings do not establish statistical superiority, human-validated faithfulness, or live-web effectiveness.

\section{\uppercase{Limitations and Threats to Validity}}
The most obvious limitation is that the corpus is synthetic rather than live. Although the mini-web is designed to emulate bulletin, news, noise, and irrelevant pages, it does not reproduce the full complexity of live web drift, adversarial content, broken markup, multilingual variability, or 
evolving information ecosystems. As a result, the reported results should be interpreted as controlled experimental findings, not direct operational performance estimates.

A second limitation is scale. The study uses 900 documents and 10 queries. That is sufficient for a structured comparative experiment, but it is small by conventional web-search standards. It limits the stability of macro averages, constrains the variety of query intent, and reduces the ability to make strong claims about generalization across broader search tasks. The study also evaluates only reranking variants within Guardian Crawler and does not include a standalone dense retriever or external neural retrieval baseline.

Third, the evaluation is domain-specific. The curated risk lexicon, query set, and synthetic document types are tailored to missing-child and trafficking-related intelligence. This is appropriate for the research question, but it also means that the strength of risk-aware reranking may not transfer directly to other domains. A risk lexicon that is highly informative here might be much noisier elsewhere.

Fourth, the mini-web generator itself can introduce bias. Choices about document-type distributions, template styles, case seed construction, and hyperlink structure inevitably shape the retrieval landscape. Even when these choices are transparent, they can privilege some signals over others. The same is true of the risk lexicon: it may emphasize some warning indicators while underrepresenting others. Because the qrel rules and risk lexicon both encode domain-related cues, their unmeasured overlap may also favor risk-aware ranking.

Fifth, the LLM-based judge is not ground truth. Judge-based grounding evaluation is useful, but it inherits model-specific brittleness, prompt sensitivity, and classification bias. Guardian Crawler partly mitigates this through separation of judge and generator roles and through lexical checks, yet the judge remains an imperfect evaluator rather than a final arbiter. Moreover, Gemini participates in corpus construction, summary generation, and judging; separate model calls provide procedural separation but not model-family independence.

Sixth, the grounded generation layer is vulnerable to case conflation. The reported failures indicate that cross-document synthesis under ambiguity remains difficult, especially when cases are lexically similar. This is not a peripheral issue; it is central to the challenge of evidence-grounded summarization in investigative contexts.

Finally, the system studies retrieval and grounded generation, not downstream operational decision-making. It does not validate search planning, intervention outcomes, or real investigative effectiveness. Those remain separate questions, although the broader missing-person analytics literature suggests several future directions for connecting retrieval outputs to spatial reasoning and search-support models \cite{ewers2024predictive,hashimoto2022agent,papic2024mobility,ruizreyes2025missing}.

% \section{Ethics, Safety, and Responsible Use}
% The application setting of this paper is sensitive. Missing-child and trafficking-related intelligence involves vulnerable populations, incomplete public information, and a real risk of harm from misinterpretation. For that reason, this paper does not present Guardian Crawler as a surveillance tool or an operational law-enforcement product. It presents it as a research platform for studying retrieval, reranking, grounded summarization, and evaluation under controlled conditions.

% Several design choices reflect this stance. First, the corpus is synthetic, which avoids direct experimentation on live vulnerable-case data. Second, the architecture is retrieval-first and evidence-linked, reducing pressure to treat generative outputs as authoritative. Third, the generation layer is bounded by retrieved evidence and explicit citations. Fourth, the evaluation layer measures faithfulness rather than assuming it. Together, these choices support transparency, reproducibility, and human oversight.

% Responsible use of systems like Guardian Crawler also requires humility about what retrieval and summarization can and cannot do. A system may surface documents or produce concise evidence-linked bullets, but it does not resolve case truth, legal responsibility, or intervention strategy. Human review remains essential, especially wherever case differentiation, ambiguity resolution, or escalation decisions are involved.

\section{\uppercase{Conclusion}}
This paper introduced Guardian Crawler, a reproducible retrieval-first testbed for controlled studies of knowledge discovery in synthetic web-like corpora motivated by safety-critical investigations. The system integrates synthetic mini-web generation, breadth-first crawling, shared HTML extraction, BM25 retrieval, LLM-based tagging, risk-aware and semantic reranking, grounded RAG-style lead generation, and explicit retrieval and automated grounding evaluation.

Within the controlled benchmark, risk-aware reranking produced the highest observed retrieval scores, increasing BM25 from macro (P@10, NDCG@10) of (0.94, 0.81) to (1.00, 0.94). The best hybrid condition reached (0.99, 0.94), and the best BM25+Semantic condition reached (0.97, 0.88). All 41 evaluable bullets passed the lexical threshold, while the automated judge rated 87.8\% as strictly supported and 90.2\% as strictly or partially supported; most identified failures involved cross-document synthesis.
%Empirically, risk-aware reranking is the strongest enhancement, raising BM25 from macro (P@10, NDCG@10) of (0.94, 0.81) to (1.00, 0.94). The best hybrid condition reaches (0.99, 0.94); the best semantic-only condition reaches (0.97, 0.88). Grounded generation is viable under tight evidence constraints, with 41/41 bullets lexically supported and 87.8\% strict and 90.2\% lenient judge-based grounding; most failures stem from cross-document synthesis, not unsupported invention.

Methodologically, Guardian Crawler provides a reproducible framework for controlled experimentation with retrieval, reranking, and automated grounding evaluation over synthetic web-like data. The current results do not establish statistical superiority, human-validated faithfulness, or transfer to live-web environments.
%Methodologically, Guardian Crawler demonstrates that retrieval, ranking, and grounded summarization in sensitive domains can be studied without collapsing the pipeline into an opaque generative system. A reproducible synthetic web, strong lexical baseline, bounded LLM roles, and explicit grounding evaluation together form a credible framework for trustworthy experimentation.

Future work includes: larger, more heterogeneous synthetic corpora; richer query sets; stronger human-in-the-loop relevance assessment; harder benchmarks with more distractors and ambiguity; better evidence aggregation to reduce case conflation; and tighter integration between retrieval outputs and downstream analytic tools for disappearance and search-support research. These are the principal requirements for determining whether the observed findings generalize beyond the present controlled testbed. %These are concrete next steps for extending retrieval-first, evidence-grounded investigative support while remaining methodologically rigorous and ethically disciplined.
\bibliographystyle{apalike}
{\small
\bibliography{example}}

@book{aggarwal2015data,
  author    = {Charu C. Aggarwal},
  title     = {Data Mining: The Textbook},
  publisher = {Springer},
  year      = {2015}
}

@book{bielska2020osint,
  author    = {A. Bielska and N. R. Kurz and Y. Baumgartner and V. Benetis},
  title     = {Open Source Intelligence Tools and Resources Handbook},
  edition   = {2020},
  publisher = {i-intelligence},
  year      = {2020}
}

@inproceedings{chen2024llmannotator,
  author    = {Rui Chen and Cheng Qin and Wenqiang Jiang and Dongwon Choi},
  title     = {Is a Large Language Model a Good Annotator for Event Extraction?},
  booktitle = {Proceedings of the Thirty-Eighth AAAI Conference on Artificial Intelligence},
  pages     = {17772--17780},
  publisher = {Association for the Advancement of Artificial Intelligence},
  year      = {2024}
}

@inproceedings{chen2022retrieval,
  author    = {Xiang Chen and Lei Li and Ningyu Zhang and Xiaoyan Liang and Shumin Deng and Chuanqi Tan and Fei Huang and Luo Si and Hsinchun Chen},
  title     = {Decoupling Knowledge from Memorization: Retrieval-Augmented Prompt Learning},
  booktitle = {Advances in Neural Information Processing Systems},
  volume    = {35},
  year      = {2022}
}

@book{croft2015search,
  author    = {W. Bruce Croft and Donald Metzler and Trevor Strohman},
  title     = {Search Engines: Information Retrieval in Practice},
  publisher = {Pearson},
  year      = {2015}
}

@misc{ewers2024predictive,
  author = {J.-H. Ewers and D. Anderson and D. Thomson},
  title  = {Predictive probability density mapping for search and rescue using an agent-based approach with sparse data},
  year   = {2024},
  note   = {Unpublished / preprint}
}

@article{floridi2019ai,
  author  = {Luciano Floridi and Josh Cowls},
  title   = {A unified framework of five principles for AI in society},
  journal = {Harvard Data Science Review},
  volume  = {1},
  number  = {1},
  year    = {2019}
}

@article{hashimoto2022agent,
  author  = {A. Hashimoto and L. Heintzman and R. Koester and N. Abaid},
  title   = {An agent-based model reveals lost person behavior based on data from wilderness search and rescue},
  journal = {Scientific Reports},
  volume  = {12},
  pages   = {5873},
  year    = {2022}
}

@article{mahadevkar2024ai,
  author  = {S. V. Mahadevkar and S. Patil and K. Kotecha and L. W. Soong and T. Choudhury},
  title   = {Exploring AI-driven approaches for unstructured document analysis and future horizons},
  journal = {Journal of Big Data},
  volume  = {11},
  pages   = {92},
  year    = {2024}
}

@book{manning2009ir,
  author    = {Christopher D. Manning and Prabhakar Raghavan and Hinrich Sch{\"u}tze},
  title     = {Introduction to Information Retrieval},
  publisher = {Cambridge University Press},
  year      = {2009}
}

@article{papic2024mobility,
  author  = {V. Papi{\'c} and A. {\v{S}}ari{\'c} Gudelj and A. Milan and M. Mili{\v{c}}evi{\'c}},
  title   = {Person mobility algorithm and geographic information system for search and rescue missions planning},
  journal = {Remote Sensing},
  volume  = {16},
  number  = {4},
  pages   = {670},
  year    = {2024}
}

@article{ratner2017snorkel,
  author  = {Alexander Ratner and Stephen H. Bach and Henry Ehrenberg and Jason Fries and Sen Wu and Christopher R{\'e}},
  title   = {Snorkel: Rapid training data creation with weak supervision},
  journal = {Proceedings of the VLDB Endowment},
  volume  = {11},
  number  = {3},
  pages   = {269--282},
  year    = {2017}
}

@article{ruizreyes2025missing,
  author  = {J. Ruiz Reyes and D. Congram and R. A. Sirbu and L. Floridi},
  title   = {Where are they? A review of statistical techniques and data analysis to support the search for missing persons and the new field of data-based disappearance analysis},
  journal = {Forensic Science International},
  volume  = {376},
  pages   = {112582},
  year    = {2025}
}
\end{document}